\documentstyle[12pt,a4]{article}
\begin{document}
\newcommand{\Ud}{\mbox{$U_3$}}
\newcommand{\rd}{\mbox{$\rho_3$}}
\newcommand{\rdk}{\mbox{$\rho_3(k)$}}
\newcommand{\rdr}{\mbox{$\rho_{3_R}$}}
\newcommand{\rdrk}{\mbox{$\rho_{3_R}(k)$}}
\newcommand{\ld}{\mbox{$\bar{\lambda}_3$}}
\newcommand{\ldr}{\mbox{$\bar{\lambda}_R$}}
\newcommand{\gdz}{\mbox{$\bar{g}_3^2$}}
\newcommand{\gdzr}{\mbox{$\bar{g}_3^2$}}
\newcommand{\ldk}{\mbox{$\bar{\lambda}_3(k)$}}
\newcommand{\ldrk}{\mbox{$\bar{\lambda}_R(k)$}}
\newcommand{\gdzk}{\mbox{$\bar{g}_3^2(k)$}}
\newcommand{\gdzrk}{\mbox{$\bar{g}_3^2(k)$}}
\newcommand{\ldkt}{\mbox{$\bar{\lambda}_3(k_T)$}}
\newcommand{\ldrkt}{\mbox{$\bar{\lambda}_R(k_T)$}}
\newcommand{\gdzkt}{\mbox{$\bar{g}_3^2(k_T)$}}
\newcommand{\gdzrkt}{\mbox{$\bar{g}_3^2(k_T)$}}
\newcommand{\be}{\begin{eqnarray}}
\newcommand{\ee}{\end{eqnarray}}
\newcommand{\zphi}{\mbox{$Z_\varphi(k)$}}
\begin{flushright}
  HD-THEP-94-31
\end{flushright}
\bigskip
\begin{center}
\huge\bf{The Strongly Interacting Electroweak
Phase \\
Transition}
\end{center}
\bigskip
\begin{center}
 B.Bergerhoff\footnote{
 e-mail: et8@ix.urz.uni-heidelberg.de}
 and C.Wetterich\footnote{
 e-mail: wetteric@post.thphys.uni-heidelberg.de} \\
 \vspace{0.5cm}
 Institut f\"ur Theoretische Physik \\
 Universit\"at Heidelberg \\
 Philosophenweg 16, D-69120 Heidelberg
\end{center}
\setcounter{footnote}{0}
\bigskip
\begin{abstract}
A quantitative discussion of nonperturbative
effects for the high temperature electroweak
phase transition is presented.
We propose a method for the computation
of the temperature dependent effective
scalar potential that takes into account 
the running of the effective gauge coupling.
Compared to perturbation theory we find a 
moderate decrease of the critical temperature
and an important change in the strength of the
first order transition.
We conclude that perturbation theory
gives a misleading picture of the dynamics
of the transition. 
\end{abstract}
\newpage
\section{Introduction}

It has recently been suggested
(\cite{A} - \cite{D})
that the electroweak high temperature
phase transition may be governed by a 
strongly interacting gauge theory.
The associated nonperturbative 
phenomena are believed to play only a modest
role if the mass of the Higgs scalar is very 
small. For realistic mass values
consistent with the experimental 
lower bounds
\cite{E},
however, the nonperturbative effects 
could dominate the quantitative and
perhaps even qualitative behaviour
of the high temperature effective
potential for the Higgs scalar.
If so, this will have important 
consequences for speculations
\cite{F}
that the observed baryon asymmetry
might have been created during the
electroweak phase transition in
the early universe.
Baryon generation within the 
standard model not only requires substantial
CP violation 
but also a sufficiently strong first
order phase transition. This is necessary
in order to guarantee a period in the
evolution of the early universe where
baryon number violating processes
are out of thermal equilibrium.
The order of the electroweak phase
transition as well as many important 
details of its cosmological dynamics
are encoded in the shape of the
high temperature effective potential.
This has recently been studied 
using different versions of resummed
perturbation theory
\cite{EWPTgeneral}.
Sizable deviations from the perturbative
results change the rates of baryon number 
violating processes or bubble formation
by many orders of magnitude and can 
therefore completely alter our
picture of the phase transition.

The importance of nonperturbative 
electroweak physics was first observed
quantitatively 
\cite{G}
in a simplified theory where all 
particles except for the Higgs scalar
were neglected.
In contrast to the perturbatively 
suggested 
\cite{3dScalarPerturb}
weak first order behaviour, the
phase transition of this model
was found to be of the second order,
with critical exponents corresponding to 
the universality class of the three
dimensional 
model. Near the transition the dimensionless 
couplings are 
strong (as dictated by the values 
of corresponding infrared fixpoints
\cite{MeanFieldAndPT}
) and perturbation theory 
becomes inadequate.
More generally, at nonvanishing
temperature $T$ the physics
of the modes with momenta
$q^2 < \left( 2\pi T \right)^2$
can always be described by an 
effective three dimensional
theory
\cite{DimensionalReduction}.
The crucial question is to what 
extent the fluctuations of these 
low momentum modes dominate the behaviour 
of the effective potential.
In the pure scalar theory an
investigation of the scale 
dependence of the effective average
potential revealed that the running
of the couplings becomes effectively three 
dimensional
(\cite{G}, \cite{H})
for scales below 
$2 \pi T$, but the difference between
the three- and the four-dimensional
running has quantitatively 
important effects only in a
narrow temperature range around
the critical temperature
where the dimensionless quartic 
scalar coupling becomes
strong.

In the electroweak standard model
one expects even stronger 
nonperturbative effects.
In the effective three dimensional
theory at scales below $2 \pi T$
the running of the gauge coupling
is not logarithmic as in four
dimensions, but gouverned by a 
power law 
\cite{A}.
In consequence, the three dimensional
``confinement scale'' was 
estimated 
between $T/10$ and $T/5$
\cite{A}.
This scale is characteristic for
the nonperturbative effects 
generated by the strong
gauge coupling and determines,
for example, the magnitude
of W-boson condensates in complete
analogy to QCD. In this work
we give a more
quantitative estimate of
the modifications of the 
high temperature effective potential 
due to the three dimensional
running of couplings.
We will perform a 
``renormalization group improvement''
for the scalar potential
$U(\varphi)$ in the effective three dimensional 
theory which is in spirit very similar
to the ``renormalization group 
improved one loop potential'' 
proposed for the zero 
temperature case by Coleman and
Weinberg
\cite{I}.
This approach accounts properly 
for the three dimensional running
of the effective couplings and
is expected to be quantitatively
very reliable as long as the
$\varphi$-dependent gauge boson
masses are sufficiently large compared
to the confinement scale.
This gives, in turn,
a lower bound on the value of the
scalar field $|\varphi|$ for which 
the potential $U(\varphi)$ is 
quantitatively under control.
For $|\varphi|$ smaller than this 
lower bound nonperturbative effects
are expected to give important 
modifications. We will give rough estimates
of the size of these effects and
demonstrate that nonperturbative physics 
dominates the phase transition for
realistic values of the mass of
the Higgs scalar.

The three dimensional physics 
is related to the infrared behaviour
of the theory. Perturbation 
theory is plagued by strong
infrared divergences in the 
presence of massless particles as
the gauge bosons for $\varphi=0$.
A suitable method to deal with these
difficulties is the average 
action $\Gamma_k$ 
\cite{AEAGeneral}
where an effective infrared 
cutoff $k$ controls the 
infrared behaviour. The 
dependence of $\Gamma_k$ on
the scale $k$ is governed by an
exact nonperturbative flow
equation
\cite{AEAEEE}.
For nonzero temperature this
flow equation accounts properly 
for the change from an effective
four dimensional running of couplings 
for $k^2 > \left( 2 \pi T \right)^2$
to the three dimensional running
for $k^2 < \left( 2 \pi T \right)^2$
\cite{G}.
We use here a simplified description
by a pure three dimensional theory
for $k < k_T$, $k_T = 2 \pi T$.
The ``initial value'' of
$\Gamma_{k_T}$ includes the 
effects of all quantum 
fluctuations except those of
the modes of the three dimensional
theory with $q^2 < k_T^2$.
It is obtained by integrating out
the $n \neq 0$ Matsubara 
frequencies, the high momentum 
modes of the $n=0$ frequency and 
all modes of the zero components
of the gauge fields $A_0$.
In particular we are interested in the 
average potential $U_k$ and we can infer 
the shape of $U_{k_T}$ from the
work of reference
\cite{J} 
(see below).
For $k \rightarrow 0$ the
average potential becomes exactly
the high temperature effective 
potential
(\cite{G},\cite{AEAEEE}).
The latter can therefore be 
computed by solving the flow 
equation for $k$ between $k_T$ and
$0$. 
As a further simplification
we leave out the fermions and
the hypercharge gauge field.\bigskip

\section{Nonperturbative flow equation}

Approximate solutions of 
the exact flow equation need
a truncation of the most general 
form of $\Gamma_k$ and we will
work with the ansatz
\be
\Gamma_k \left[ \varphi, A_\mu \right] =
\int d^dx \left( U_k(\rho) +
Z_{\varphi,k} | D_\mu \varphi |^2 +
\frac{1}{4} Z_{F,k} F_{\mu\nu} F^{\mu\nu}
\right) .
\label{1}
\ee
Here $\rho = \varphi^\dagger 
\varphi$ and group indices are omitted.
The form of the average potential
$U_k(\rho)$ is left arbitrary
and has to be determined by
solving the flow equation.
For the Abelian Higgs model the 
evolution equation for the 
average potential was computed
in the approximation (\ref{1})
in reference
\cite{AEAAHMOld}.
Inserting the appropriate
$SU(2)$ group factors
we obtain in the Landau
gauge ($\alpha=0$), with 
$t=\ln k$
\be
\frac{\partial}{\partial t} U_k(\rho) &=& 
\frac{1}{2} \int\!\! 
\frac{d^d q}{\left( 2 \pi \right)^d}
\frac{\partial}{\partial t}
\left( 3 \left( d-1 \right)  
\ln \left( q^2+k^2+ m_B^2 \right) + \right. \nonumber \\
& & \left. + \ln \left( q^2+k^2 + m_1^2 \right) +  
3 \ln \left( q^2+k^2 + m_2^2 \right) \right) 
\label{2}
\ee
where the mass terms read
\be
m_B^2 = \frac{1}{2} Z_{\varphi,k} \bar{g}^2 \rho \ ; \
m_1^2 = \left( U_k'(\rho) +
2 \rho U_k''(\rho) \right) / Z_{\varphi,k} \ ; \
m_2^2 = U_k'(\rho) / Z_{\varphi,k} .
\label{3}
\ee
We have used here a masslike 
infrared cutoff $R_k = Z_k k^2$
but more general cutoff functions
$R_k(q)$ may be employed.
The partial derivative 
$\frac{\partial}{\partial t}$ on the
right hand side of (\ref{2}) is
meant to act only on $R_k$ and
we omit contributions arising 
from the wave function renormalization
$Z_k$ in $R_k$. Primes denote 
derivatives with respect to $\rho$.

This flow equation can be used\footnote{
The ultraviolet divergence on the 
right hand side of equation (\ref{2}) is
particular to the use of a masslike
infrared cutoff. For $d < 4$ it
concerns only an irrelevant constant
in $U$ and is absent for 
$\partial U'/\partial t$.}
in 
arbitrary dimensions $d$ and constitutes 
a nonlinear partial differential
equation for the dependence
of $U$ on the two variables
$k$ and $\rho$.
In our case it holds for the
three dimensional potential
$U_3$ and a correspondingly 
normalized scalar field 
$\rho_3$. They are related to
the usual four dimensional 
quantities by $U_3 = U_4 / T$ ,
$\rho_3 = \rho_4 / T$.
The gauge coupling in (\ref{3})
stands for the three dimensional
running renormalized gauge coupling
$\bar{g}_3^2(k)$. Its value
at the scale $k_T$ is given by
\be
\bar{g}_3^2(k_T) = g_4^2(k_T) T
\left( 1 - \frac{g_4^2(k_T) T}
{24 \pi m_D} \right)
\label{4}
\ee
where
\be
m_D^2 = \frac{5}{6} g_4^2(k_T) T^2 
\label{neu1}
\ee
and the effects of integrating
out the $A_0$ mode have been included
in lowest order
\cite{J}.
The evolution equation for the 
running gauge coupling in the pure 
Yang-Mills theory has been computed in 
reference 
\cite{A}.
We use here mainly the lowest order 
result 
\be
\frac{\partial}{\partial t} \gdzr = \beta_{g^2} =
- \frac{23 \tau}{24 \pi} {\bar{g}_{3}}^4(k) k^{-1} 
\label{5} 
\ee
where the deviation of $\tau$ 
from one accounts for the 
small
contributions of scalar fluctuations
which remain to be computed\footnote{
For a suitable choice of wave function
renormalization constants in the infrared 
cutoff for the gauge bosons the lowest 
order result becomes independent of
the gauge parameter $\alpha$ and can therefore be
used for the Landau gauge employed 
in this paper.}.
Furthermore, we need the anomalous dimension
of the scalar field. For
our purpose it can be approximated by
\cite{AEAAHMOld}
\be
\eta_\varphi = - \frac{\partial \ln Z_\varphi}{\partial t}
 = - \frac{1}{4\pi} \gdzrk k^{-1} . 
\label{6}
\ee
\bigskip

\section{Running quartic coupling}

A convenient quantity for an investigation
of the effective potential is the $\rho$-dependent
quartic coupling 
\be
\bar{\lambda}_{3,k}(\rho)
= U_k''(\rho) 
= \frac{\partial^2 U_{3,k}}{\partial \rd^2} .
\label{7}
\ee
Knowing for $k=0$ the function 
$\bar{\lambda}_3(\rho) = 
\bar{\lambda}_{3,0}(\rho)$,
the high temperature effective 
potential $U(\rho) = U_0(\rho)$
can be reconstructed by integration
and translation to a four
dimensional normalization.
One of the two integration
constants is irrelevant and 
the other (the mass term 
linear in $\rho$) can be found by adapting
$U(\rho)$ to the perturbative 
result for large $\rho$ where
the three dimensional
running of the couplings is irrelevant.
In fact, for large $\rho$ 
such that $m_B^2(\rho) > k_T^2$ 
one may use the one loop potential 
\cite{HTPTOneLoop}
\be
\tilde{U}_3(\rd) = -\mu^2(T) \rd \!+\!
\frac{1}{2} \!\left( \ld + \Delta \ld
\right) \!\rd^2 \!-\! \frac{1}{12 \pi} \!\left(
6 m_B^3 \!+\! 3 m_E^3 \!+\! \bar{m}_1^3 \!+\! 3 \bar{m}_2^3 \right)
\label{8}
\ee
with $\bar{m}_1^2 = 3 \ld \rd - \mu^2(T)$,
$\bar{m}_2^2 =  \ld \rd - \mu^2(T)$,
$Z_\varphi = 1$. 
Here 
\be
\Delta \ld = \frac{3 \bar{g}_3^4}{
64 \pi^2 T} \left( 1 +
\frac{\sqrt{6}+\sqrt{2}}{8} 
\frac{M_h^3}{M_w^3} \right)
\label{8.1}
\ee
with $M_h$ and $M_w$ the 
masses of the Higgs scalar and the
gauge boson respectively
is chosen such that
$\tilde{U}_3''(8 \pi^2 T^2 / \gdzkt) = \ld$.
Two loop effects and
corrections from integrating out 
$A_0$ may be included 
\cite{J}.
We use here the mass term $\mu^2(T)$
as given by
\be
\frac{\mu^2(T)}{T^2} = 
\frac{1}{2} \frac{M_h^2}{T^2} -
\frac{3 \pi}{4} \alpha_w c -
\frac{\pi}{4} \alpha_w \frac{M_h^2}{M_w^2}
\label{9}
\ee
where
$\alpha_w = g_4^2(k_T)/4\pi$,
the weak fine structure constant 
at the scale $k_T$.
Here $c = 1 - \frac{m_D}{\pi T} + \Delta c$ includes
effects from integrating out the 
$A_0$ mode. We use $\Delta c = 0$ but higher loop 
corrections or the effects from including
the quark fluctuations can be accounted
for by an appropriate nonvanishing
$\Delta c$. For example including
the top quark and using $6$  quark flavours for
the Debye mass yields
$ \Delta c = \frac{2 M_{top}^2}{3 M_w^2} + \frac{\sqrt{5}-\sqrt{11}}
{\sqrt{6}} \frac{g_4}{\pi}$
where the second term accounts for the 
modification of $m_D$.
The initial conditions for the flow 
equation described below will be expressed
for fixed $\mu^2(T)/T^2$ and
$M_h^2/M_w^2$. In consequence, 
$\Delta c \neq 0$ results in a 
simple rescaling of temperature and the
scalar field according to
\be
T_{mod}^{-2} &=& T^{-2} +
\frac{3 \pi \alpha_w}{2 M_h^2} \Delta c \nonumber \\
\varphi_{mod}/\varphi &=& T_{mod}/T .
\label{new11}
\ee
Here $T_{mod}$ is appropriate for
$\Delta c \neq 0$ whereas $T$ 
represents the scaling used in this
work for $\Delta c = 0$.
The above simple rescaling property
is very useful for a quantitative
comparison with authors using
different prescriptions for
the effective three dimensional
theory.

The evolution equation for the $k$-dependence 
of $\bar{\lambda}_3(\rho_3)$ can
be infered from (\ref{2}) by 
differentiating twice with respect to
$\rho_3$ and reads
\be
\frac{\partial \bar{\lambda}_{3,k}
(\rd)}{\partial t}  = \frac{3}{32 \pi} \!\left(
\frac{Z_{\varphi,k}^2 \bar{g}_{3,k}^4 k^2}
{\left( k^2+m_B^2 \right)^{3/2}} \!+\!
\frac{6 Z_{\varphi,k}^{-2} \bar{\lambda}_{3,k}^2(\rd) k^2}
{\left( k^2+m_1^2 \right)^{3/2}} \!+\! 
\frac{2 Z_{\varphi,k}^{-2} \bar{\lambda}_{3,k}^2(\rd) k^2}
{\left( k^2+m_2^2 \right)^{3/2}} \right) 
\label{10}
\ee
where we have neglected terms 
$\propto U_k^{(3)}(\rd)$ and 
$U_k^{(4)}(\rd)$.
We propose in this paper
an approximate solution of the flow equation 
for $\bar{\lambda}_3(\rd)$ for $k=0$. 
It is based on the observation that
the \rd-dependent mass terms 
$m_B^2$, $m_1^2$, and $m_2^2$ act in
equation (\ref{2}) as independent infrared 
cutoffs in just the same way as $k^2$.
A variation of $m^2$ for $k^2=0$ is
roughly equivalent to a variation
of $k^2$ at $m^2=0$.
We use this observation to translate the flow equation 
(\ref{10}) into a renormalization 
group equation for $\ld(\rd)$ at 
$k=0$: In equation (\ref{10}) we
replace $\frac{\partial}{\partial t}$ by
$\frac{\partial}{\partial t'} = 
m_B \frac{\partial}{\partial m_B}$ and the factors
$k^2 \left(k^2+m^2 \right)^{-3/2}$ by
$m^{-1}$.
In order to see that our simple
prescription indeed corresponds to an
approximate solution of the flow equation
(\ref{10}) we omit for a moment the two last 
terms in (\ref{10}) which 
arise from the scalar fluctuations.
Writing 
\be
\frac{\partial}{\partial t} 
\left( m_B \frac{\partial}{\partial m_B}
\bar{\lambda}_{3,k}(\rd) \right) =
\frac{3}{32\pi} Z_\varphi^2(k) \bar{g}_3^4(k)
m_B^2 \frac{\partial}{\partial t}
\left( k^2+m_B^2 \right)^{-3/2}
\label{11}
\ee
we observe that the right hand side is
suppressed both for $k^2 \gg m_B^2$ and 
$k^2 \ll m_B^2$. Since the evolution
of the quantity $m_B \frac{\partial}
{\partial m_B} \ld$ is dominated 
by a narrow interval $k \simeq m_B$,
we may approximate on the right hand side
of equation (\ref{11}) \gdzk\ and $Z_\varphi(k)$
by $\bar{g}_3^2(m_B)$ and $Z_\varphi(m_B)$,
and similar in the relation (\ref{3}) between
$m_B^2$ and \rd. Equation (\ref{11}) is then
easily integrated between $k=0$ and 
$k=k_T$ to give
\be
m_B \frac{\partial}{\partial m_B}
\bar{\lambda}_{3,0}(\rd) &=&
\frac{3}{32\pi} Z_\varphi^2(m_B)
\bar{g}_3^4(m_B) m_B^{-1} +
m_B \frac{\partial}{\partial m_B}
\bar{\lambda}_{3,k_T}(\rd) - \nonumber \\
& & - 
\frac{3}{32\pi} Z_\varphi^2(m_B)
\bar{g}_3^4(m_B) 
\frac{m_B^2}{\left(
k_T^2+m_B^2 \right)^{3/2}} .
\label{12}
\ee
The sum of the last two terms is
small compared to the first term\footnote{
This statement is immediately apparent
for $m_B \ll k_T$ where both terms are
seperately small. We expect it to hold 
also for $m_B \sim k_T$, but a proof
needs a detailed understanding of the 
transition region between the 
effective three dimensional and
the effective four dimensional
theory.}
which coincides exactly with our prescription.
This type of arguments can be generalized
for the scalar fluctuations.
In summary, we can now work with a
new effective infrared cutoff $k'=m_B$
which is a function of \rd\ 
(we omit the prime on $k$ in the 
following),
\be
k^2 = m_B^2 = \frac{1}{2} Z_\varphi(k) \gdzk \rd .
\label{13}
\ee
This procedure transforms equation (\ref{10})
into a simple differential equation
for $\bar{\lambda}_3(\rd) = 
\bar{\lambda}_3(k(\rd))$. In terms
of the renormalized coupling
\be
\ldrk = Z_\varphi^{-2}(k) \ldk 
\label{14}
\ee
it reads
\be
\frac{\partial}{\partial t} \ldrk =
\frac{3}{32 \pi k} \left( 
\bar{g}_3^4(k) + 
\left( \sqrt{6} + \sqrt{2} \right) \bar{\lambda}_R^{3/2}(k)
\bar{g}_3(k) \right) 
- \frac{1}{2 \pi k} \gdzrk \ldrk . 
\label{15}
\ee
For the terms $\propto \frac{1}{m_1}$
and $\propto \frac{1}{m_2}$ from 
equation (\ref{12})
we have approximated
in (\ref{3}) $U'(\rd) \simeq
\rd \bar{\lambda}_R(\rd)$ which amounts
to neglecting the mass term. For negative
\ldrk\ our approximation does
not describe properly the effect of the 
scalar fluctuations. Since their 
contribution is small in this 
region we simply omit the terms 
$\propto \bar{\lambda}_R^{3/2}$
once \ldrk\ becomes negative.

The renormalization group equation
(\ref{15}) for the \rd-dependence of
$U''$ is the central equation of this 
work. Except for the last term
arising from the anomalous dimension it 
can be directly obtained by taking
appropriate derivatives of the one
loop formula (\ref{8}), 
treating mass ratios such as 
$m_B/m_1$ as $k$-independent and 
replacing at the end the couplings
\gdz\ and \ld\ by running couplings
evaluated at the scale $k$.
For $Z_\varphi=1$ and \gdz, \ld\ independent 
of $k$ equation (\ref{15}) reproduces
exactly the one loop result.
Our renormalization group equation enables
us to include the effects of running 
couplings and the anomalous dimension.
Combining equations 
(\ref{5}), (\ref{7}),
(\ref{13}), (\ref{14}) and (\ref{15}) we can
compute the \rd-dependence of the
high temperature effective potential 
by a solution of the flow equation.
The initial values for this solution
are set by equations 
(\ref{4}), (\ref{9}) and 
\be
\bar{\lambda}_3(k_T) = 
\frac{1}{4} g_4^2(k_T) T 
\frac{M_h^2}{M_w^2} -
\frac{3 g_4^4(k_T) T^2}{64 \pi m_D} +
\Delta \ld 
\label{16}
\ee
where the last value is extracted 
from the validity of the one loop
potential (\ref{8}) at $k_T$.\bigskip

\section{Analytical approximations}

Before turning to the numerical 
solution of the flow equations we
display some analytical results in appropriate
approximations.

First, from (\ref{5}) one easily finds that the running gauge
coupling is given by
\be
\frac{1}{\gdzrk} = \frac{1}{\gdzrkt} +
\frac{23 \tau}{24 \pi} \left( \frac{1}{k_T} - 
\frac{1}{k} \right) .
\label{17}
\ee
Thus, \gdzrk\ diverges as $k$ approaches 
$k_\infty$ with 
\be
k_\infty = \left( \frac{1}{k_T} +
\frac{24 \pi}{23 \tau \gdzrkt} 
\right)^{-1} .
\label{18}
\ee
This reflects the well known fact that the $SU(2)$ Higgs
model is confining in 3 dimensions. 
From (\ref{6}) we can compute $Z_\varphi(k)$ 
and find
\be
Z_\varphi(k) = \left( 
\frac{\gdzrk}{\gdzrkt} \right)^{-\frac
{6}{23 \tau}} ,
\label{19}
\ee
where we set $Z_\varphi(k_T) = 1$. Using (\ref{19})
we can explicitly relate the renormalized to the 
unrenormalized couplings.
Also, \rdk\ is given by
\be
\rdk = \frac{k^2}{2 \pi \alpha_w T}
\left( 1 + \frac{23 \tau\alpha_w}{12 \pi}
\left(1-\frac{2\pi T}{k} \right) 
\right)^{1-\frac{6}{23\tau}}  .
\label{20}
\ee

For small \ldrk\ we can neglect the term $\propto 
\bar{\lambda}_R^{3/2}$ on the right hand 
side of (\ref{15}) 
and solve this equation analytically.
One easily recovers the 1-loop result if
one sets $ \eta_\varphi  = 0$
and $\gdz = const.$. 
We can also find a solution
if we include $\eta_\varphi$ and
the running of the gauge
coupling.
Defining 
\be
R(k) = \frac{\bar{\lambda}_R(k)}{\gdzrk}
\label{21}
\ee
the flow equation reads
\be
\frac{dR}{d\ln\gdzr} = 
- \frac{9}{92 \tau} -\left(
1 - \frac{12}{23 \tau} \right) R .
\label{22}
\ee
This is easily solved for $R$ and one finds
\be
\bar{\lambda}_R(k) = 
\left( \frac{\gdzrk}{\gdzrkt} \right)^{
\frac{12}{23 \tau}}
\bar{\lambda}_R(k_T) -
\frac{9}{92 \tau - 48} \gdzrk 
\left( 1 - \left( \frac{\gdzrk}{\gdzrkt} 
\right)^{\frac{12}{23 \tau}-1} \right) .
\label{23}
\ee
The main 
changes as compared to the 1-loop calculations
can be understood
from the corresponding differential equations:
The inclusion of $\eta_\varphi$ lowers the scale
at which $U''$ changes
sign, whereas the running of the gauge coupling
acts the opposite way. Thus at a given $k$,
corresponding to a given \rd, the running of 
\gdzr\ makes the potential bend up less
than the 1-loop calculation predicts.
The inclusion of the neglected terms
$\propto \bar{\lambda}_R^{3/2}$ will
have the same effect as soon as they are 
not small, i.e.\,\,for larger Higgs masses.

In fig.\,1 we have depicted the quartic coupling 
$U''(\rho)$ (in a four dimensional 
normalization) as a function of
$\Phi = \rho_4^{1/2}$ for $M_h = 35 \mbox{\,\,GeV}$.
The corrections due to the running
gauge coupling and the anomalous dimension are small
for $\Phi > 100 \mbox{\,\,GeV}$. There are 
sizable modifications in the region
$40 \mbox{\,\,GeV} < \Phi < 60 \mbox{\,\,GeV}$ and large
corrections for $\Phi < 40 \mbox{\,\,GeV}$.
Translated to the effective 
potential the running of
\gdz\ should strengthen the phase transition and lower 
the critical temperature. This is what one would
naively expect, since the first order character of the transition 
is due to the gauge-boson loops. 
Thus, enhancing the coupling should give
a transition more strongly first
order. We should note, however, that the anomalous 
dimension has the opposite effect. To estimate the
precise effects of these modifications
on the shape of the potential we have to 
proceed to a numerical integration of equation 
(\ref{15}).\bigskip

\section{Numerical solution of the flow equation}

In order to make the numerical solution as
easy as possible, we have transformed the 
system of differential equations (\ref{7})
and (\ref{15}) into a third order differential
equation for \Ud. Since equation (\ref{20})
gives \rd\ 
in terms of $k$ and is not easily inverted 
analytically, we use $k$ as the integration variable.
This yields
\be
\frac{\partial^3 \Ud}{\partial k^3} =
A^2 Z_\varphi^2 \beta^{(1)} + 
3 \frac{B}{A} \frac{\partial^2 \Ud}
{\partial k^2} - 
\left( 3 \frac{B^2}{A^2} - \frac{C}{A}
\right) \frac{\partial \Ud}{\partial k}
\label{25}
\ee
where
\be
\beta^{(1)} =    
\frac{3}{32 \pi k^2} \left( 
\bar{g}_3^4(k) + 
\left( \sqrt{6} + \sqrt{2} \right) \bar{\lambda}_R^{3/2}(k)
\bar{g}_3(k) \right) \nonumber \\
A(k) = \frac{\partial \rd}{\partial k} \  ; \ 
B(k) = \frac{\partial^2 \rd}{\partial k^2} \  ; \ 
C(k) = \frac{\partial^3 \rd}{\partial k^3} .
\label{26}
\ee
As initial conditions we use
\be
\hspace{-2.5cm}\frac{\partial^2 \Ud}{\partial k^2}(k_T) =
A^2(k_T) 
\ldkt
 + B(k_T) 
\frac{\partial \tilde{U}_3}{\partial \rd}(k_T) \nonumber \\
\frac{\partial \Ud}{\partial k}(k_T) =
A(k_T) 
\frac{\partial \tilde{U}_3}{\partial \rd}(k_T) \ ;  \
U_3(k_T) = \tilde{U}_3(k_T) \ ; \ 
Z_\varphi(k_T) = 1 .
\label{27}
\ee
$\tilde{U}_3$ represents the 1-loop
potential (\ref{8}) such that
\be
\frac{\partial \tilde{U}_3}{\partial \rd}(k_T)\,= 
-\mu^2(T) \!+\!
\left(\ld \!+\! \Delta \ld \right)\! 
\frac{8 \pi^2 T^2}{\gdzkt} \!-\!
\frac{9}{12} \!T\! \left(
\gdzkt \!+\! \left(
\sqrt{6}\!+\!\sqrt{2} \right)\!
\frac{\bar{\lambda}_3^{3/2}}{\bar{g}_3(k_T)}
\right) .
\label{28}
\ee
A similar set of equations relates the
derivatives of the potential with respect to
\rd\ to the solution of (\ref{25}).
We integrate equation
(\ref{25}) with the help of the Runge-Kutta
method and use for all numerical work 
$g_4 = 2/3$, 
$M_w = 80.6 \mbox{\,\,GeV}$, and
$\tau = 1$. 

In fig.\,2 we show the effective potential as
obtained with our method for different
temperatures and for masses of the Higgs scalar 
of $35$ and $80 \mbox{\,\,GeV}$. In all the plots we use 
four dimensional quantities and plot
$\Delta U = U(\Phi)-U(0)$. 
For $M_h = 35 \mbox{\,\,GeV}$ (fig.\,2a) the first
two curves correspond to the critical 
temperature as given by the one loop approximation
($T=97.50\mbox{\,\,GeV}$) and the critical temperature
obtained from our renormalization group
improved approach ($T=95.85\mbox{\,\,GeV}$).
Even though the critical temperature 
is not changed by much, the shape of the 
potential at a given temperature 
varies considerably. At the one loop
critical temperature there remains not
even a local minimum once renormalization
group effects are included! 
We also find important quantitative 
differences for the size of the
barrier at the critical temperature 
in both approaches. 
This effect becomes larger for 
higher values of the Higgs mass.
For $M_h = 80 \mbox{\,\,GeV}$ (fig.\,2b) the critical temperatures
($185.23\mbox{\,\,GeV}$ for one loop, 
$184.25\mbox{\,\,GeV}$ for RG-improvement)
remain similar, but the shape of the 
potential differs strongly
(even though this is difficult 
to see on the scale we use in 
fig.\,2b). We have also indicated by a 
square the value of $\Phi$ where
the dimensionless gauge coupling 
$g^2(k) = \gdzk/k$ becomes as
large as $2\pi$, and similarly, by the end 
of the solid line, where it reaches 
$4\pi$. These points may be considered
as an estimate for the limit of validity
of our approximations.
To the left of this region we have to
deal with a strongly interacting
gauge theory. Straightforward
use of our approach in this region
would yield the potential as indicated by
the diamonds. We observe that our
renormalization group improved
effective potential can be formally
extended to $\Phi=0$: The confinement
scale $k_\infty$ corresponds to 
$\rho=0$ since by equation
(\ref{13}) $\rho_3 \propto \bar{g}^{-2}_3(k) k^2$ 
and $\bar{g}^2_3(k_\infty) = \infty$.
This explains the different qualitative
behaviour as compared to \cite{C}.

In order to demonstrate that 
$g^2 \sim 2\pi - 4\pi$ indeed corresponds
to the onset of the strong coupling regime
we notice that $g^2(k)=2\pi$ is reached
at a scale $k \sim 1.5 k_\infty$, which is
already very close to the confinement 
scale (cf.\,\,eq.\,\,(\ref{17})).
The anomalous dimension $\eta_\varphi$
reaches the value $-1$ for 
$g^2(k) \sim 4\pi$ (cf.\,\,eq.\,\,(\ref{6})).
Instead of the lowest order estimate for
the running of \gdz\ (\ref{5}) we may also use the 
improved nonperturbative estimate 
of reference  \cite{A},
\be
\frac{\partial \gdzk}{\partial t} =
- \frac{23}{24 \pi k} \left( \frac{
1}{1-\frac{21}{48\pi} \gdzk k^{-1}} \right)
\bar{g}_3^4(k) .
\label{30}
\ee
Here the $\beta$-function diverges\footnote{
This is, of course, not a physical effect;
see the discussion in \cite{A}} for
$g^2(k) = \frac{48}{21}\pi$, which is 
a value close to $2\pi$.
For the $\beta$-function (\ref{30}) the
confinement scale turns out higher and corresponds 
now to a nonzero value
of $\Phi$. We compare the results of 
the $\beta$-functions (\ref{5}) and 
(\ref{30}) for the effective potential in 
fig.\,3, where we also display the one
loop potential at the same temperature.
The diamonds correspond to (\ref{30}) 
and this curve ends once $\beta_{g^2}$
diverges. For ease of comparison we
have shifted all three curves to conicide 
at $\Phi=100 \mbox{\,\,GeV}$. The temperature chosen
is the critical temperature for $M_h=80 \mbox{\,\,GeV}$
as obtained by the renormalization group
improved potential with the lowest order
$\beta$-function (\ref{5}).
We emphasize that strong interaction phenomena
cover here the whole region between the two 
minima. Any perturbative estimate 
of details of the potential as needed for
calculations of bubble nucleation and for 
a treatment of the cosmological 
dynamics of the phase transition seems
highly unreliable for Higgs masses of about 
$80 \mbox{\,\,GeV}$ or higher. 

At this point we should mention that the
renormalization group improvement has 
already extended considerably the range 
of $\Phi$ where a reliable 
computation is available: In the 
standard perturbative loop calculations
the $\varphi$-dependence of the 
effective gauge coupling is usually not
taken into account. Since the 
gauge-boson fluctuations are crucial
for creating the barrier responsible for
the first order transition, a change 
of \gdz\ by a factor of two as 
compared to the perturbative value 
(which in our language corresponds
to $\bar{g}_3^2(k_T)$) seems to be
at the limit of what is tolerable
for a 
perturbative estimate
to be quantitatively
correct. From (\ref{17}) we find that
$\gdzk/\gdzkt=2$ corresponds to a
value of $\Phi=0.55 T$ (we
have indicated this value by a 
cross on the curves in fig.\,2).
In contrast, the value where
$g^2(k) = 2\pi$ (squares in 
fig.\,2) corresponds to
$\Phi_{np} = 0.28 T$ which is smaller
by a factor of two. Inspection
of figures 1 and 2a suggests that even the range 
$\Phi > 0.55 T$ for the validity
of the loop expansion is overestimated
and a more conservative 
estimate amounts to
$\Phi \stackrel{>}{\sim} T$.\bigskip

\section{Strong electroweak interactions}

We finally turn to the region of strong
interactions to the left of the solid
line in fig.\,2. We first observe that our computation 
of the potential has effectively included 
only the quantum fluctuations with 
$q^2>k_\infty^2$ since even for $\Phi=0$ the infrared
cutoff is given by $k_\infty$.
In consequence, the potential we
have computed and shown in figures 2 and 3
corresponds to the average potential
$U_{k_\infty}$ rather than to
the effective potential $U_0$. 
For $\Phi$ sufficiently large, i.e.\,\,for
$\Phi>\Phi_{np}$ to
the right of the squares in fig.\,2, the 
difference between $U_0$ and 
$U_{k_\infty}$ should be small 
since the contribution of the low momentum 
modes is suppressed
by an effective infrared cutoff
$m_B \gg k_\infty$. This is not
true anymore in the region of
small $\Phi$ where the low momentum fluctuations are 
expected to produce large nonperturbative
effects. In particular, we notice that the 
ansatz (\ref{1}) becomes insufficient: 
At the scale $k_\infty$ 
the gauge boson kinetic term 
$\propto Z_F(k_\infty)$ vanishes 
$Z_F(k)$
and becomes negative 
\cite{A}
for $k<k_\infty$. This is a clear indication
that for $k=0$ the minimum of the 
Euclidian effective action does not
occur for $F_{\mu\nu}F^{\mu\nu}=0$ but
rather for a nonzero value of $F^2$
\cite{K}.
One expects W-boson condensates in close analogy
to the gluon condensates in QCD. The best way to visualize
these effects in the context of the 
effective average action is perhaps 
the introduction of a composite 
scalar field $\Theta$ for
the operator $\frac{1}{4} F_{\mu\nu}F^{\mu\nu}$.
This can be done at a scale in the
vincinity of (somewhat above) $k_\infty$
according to the general formalism
proposed in 
\cite{L}.
The average scalar potential is then generalized
to a potential depending on
two scalar degrees of freedom,
$U_k(\Phi,\Theta)$. For
$k>k_\infty$ the minimum of
$U_k(\Phi,\Theta)$ occurs\footnote{
Strictly speaking $<\!\!\Theta\!\!>$ may take
a nonzero value depending on its 
precise definition. For the purpose 
of this qualitative discussion we 
shall speak of $<\!\!\Theta\!\!>=0$ if 
the effect of $<\!\!\Theta\!\!>$ is small.} at 
$<\!\!\Theta\!\!>=0$
whereas for $k\rightarrow 0$ the minimum value 
$<\!\!\Theta\!\!>$ will be a nonvanishing 
function of $\Phi$ with  
$<\!\!\Theta\!\!>(\Phi) > 0$ for $\Phi$ smaller
than some critical $\Phi_{cr}$. 
In view of fig.\,3 we may roughly associate
the critical $\Phi_{cr}$ for the onset 
of condensation phenomena with 
$\Phi_{np}$, denoted by the squares in fig\,2. 
The W-boson condensate will lower the value of the
effective potential
$U(\Phi)=U(\Phi,<\!\!\Theta\!\!>(\Phi))$ in the region of
small $\Phi$ as compared to the
computed $U_{k_\infty}(\Phi,0)$
\cite{B}.
We will present in the following some quantitative 
but rough estimates how this affects our
picture of the phase transition.

Let us try to estimate the difference
\be
\delta_1 = U_{k_\infty}(0,0)-
U_0(0,<\!\!\Theta\!\!>(0))
\label{31}
\ee
which measures how far the W-boson
condensation lowers the potential at the 
origin. Since for $\Phi=0$ the only
relevant scale for the
condensate is the confinement scale
$k_\infty$, we can infer from simple
dimensional analysis
\be
\Delta U_3 = K k_\infty^3
\label{32}
\ee
with $K$ a constant of order one. Since the 
running of the gauge coupling in 
three dimensions obeys a power law
(in contrast to the four dimensional
logarithms), the scale $k_\infty$ is
determined relatively precisely once the 
complete beta-
function is known.
In the relevant region the gauge 
coupling is large and there is no apparent
small dimensionless parameter in the problem
and we will therefore
use $K=1$. With $k_\infty = 0.13 T$ we infer
for $\delta_1$
\be
\delta_1 = 2.2 \times 10^{-3} K T^4 .
\label{33}
\ee
We have depicted $\delta_1$ in figure 2 and also 
show the potential for the 
``nonperturbative critical temperature''
corresponding to 
$T=94.70\mbox{\,\,GeV} (173.74\mbox{\,\,GeV})$ for
$M_h = 35 \mbox{\,\,GeV} (80 \mbox{\,\,GeV})$. For
$M_h = 35 \mbox{\,\,GeV}$ the change in the critical
temperature is not very large, but the
size of the barrier as obtained by 
extrapolating the potential from
$\Phi=\Phi_{np}$ to the value
$\Delta U = -\delta_1$ for $\Phi=0$
is considerably enhanced by the 
W-boson condensation. 
For $M_h = 80 \mbox{\,\,GeV}$ this effect is
even more dramatic, and also the
critical temperature is lowered
considerably. A combination of
our renormalization group improved
potential for $\Phi>\Phi_{np}$ and
other nonperturbative estimates of
the critical temperatures using methods
sensible to condensation phenomena 
- for example lattice calculations - 
can be used to determine the proportionality 
factor $K$. For this purpose we
plot in figure 4 the critical temperature
as a function of the Higgs-mass for various 
values of $K$.
Here we have also included the (properly
rescaled) results of recent lattice
studies for $M_h = 35\mbox{\,\,GeV}$
\cite{KRS}
and
$80\mbox{\,\,GeV}$
\cite{FKRS}
respectively. These values seem to
support the general picture, yielding 
$K$ of the order of $10$. Conclusive
evidence surely needs more data
however.

It is also interesting to obtain an
estimate for a lower
bound on the critical temperature.
For this purpose we use
\be
\delta_2 = U_{k_\infty}(\Phi_{np},0)-
U_0(0,<\!\!\Theta\!\!>(0)) = K' k_{np}^3 T \simeq
8 \times 10^{-3} K' T^4 .
\label{34}
\ee
Since, as argued above, $k_{np}$ denotes the scale
for the onset of strong interaction phenomena
it is difficult to conceive that the 
proportionality constant can exceed one by much.
We display $\delta_2$ for 
$K' = 1$ together with the potential 
for the corresponding critical temperature 
in fig.\,2. The ``lower bounds'' on the
critical temperature are
\be
T_{cr}&>&92.54\mbox{\,\,GeV}  \mbox{\ for\ }
 M_h = 35 \mbox{\,\,GeV} \nonumber\\
T_{cr}&>&159.14\mbox{\,\,GeV}  \mbox{\ for\ }
 M_h = 80 \mbox{\,\,GeV} 
\label{35}
\ee
and somewhat below these values for
$K' > 1$.
It is apparent from fig.\,2 that the
lower bounds on $T_{cr}$ also correspond
to upper bounds on the surface under the 
barrier which determines the strength of
the first order phase transition.
We observe that the use of the improved
$\beta$-function (\ref{30}) enhances 
$k_\infty$ by a factor of two and 
leads to the guess $K=8$. This yields 
$T_{cr}$ in the vincinity of
the ``bound'' (\ref{35}),
and seems to be favoured by
the lattice results available.

Our estimates can be extended to
larger values of the Higgs mass which are
up to now difficult to access by alternative
methods. In fig.\,5 we plot the effective potential
for $M_h = 140 \mbox{\,\,GeV}$ and $200 \mbox{\,\,GeV}$ for three values
of the temperature: The upper curve corresponds
to the critical temperature as determined from 
$U_{k_\infty}$ in the absence of condensation 
phenomena, the middle curve shows our estimate 
using $K=1$
for the critical temperature including W-boson 
condensation, and the lower curve corresponds to the 
``lower bound'' given above. For 
$M_h = 200 \mbox{\,\,GeV}$ we notice that $\Phi_{np}$ 
(again indicated by a square) almost 
approaches the minimum of $U$ for
$T_{cr}=318.9\mbox{\,\,GeV}$.
This may have important consequences
for the dynamics of the phase transition: 
It is conceivable that there is almost no
barrier between the minima at 
$\Phi=0$ and $\Phi \neq 0$, and that
the barrier is even completely absent 
for $M_h \stackrel{>}{\sim} 200 \mbox{\,\,GeV}$.
This could be interpreted as a
change from a first order
transition to an analytical crossover
for very large Higgs masses
\cite{A}.
Even though there is at presence no
evidence for an analytical crossover,
this possibility cannot be
excluded for $M_h \stackrel{>}{\sim} 200 \mbox{\,\,GeV}$!
\bigskip

\section{Conclusions}

In conclusion we have presented
here a renormalization group improved estimate 
of the high temperature effective potential
which determines the dynamics
of the electroweak phase transition in
the early universe. It is
based on an approximative solution of
an exact nonperturbative flow equation 
and includes properly 
several effects not accounted for in
the presently available results
from the loop expansion.
This concerns, in particular, the running
of the effective three dimensional
gauge coupling and the anomalous dimension
of the scalar field.
We believe our estimates to be 
quantitatively accurate for
large enough values of the scalar field,
$\Phi > \Phi_{np}$.
In this region of field space, the differences as 
compared to the loop expansion are already considerable even 
for a mass of the Higgs scalar as low
as $35 \mbox{\,\,GeV}$, being further enhanced
for large scalar masses. We argue that 
nonperturbative effects not included in the present calculation
further lower the critical temperature. They also
enhance the barrier characteristic for the
strength of the first order phase transition for
low and moderate scalar masses
($M_h \stackrel{<}{\sim} 100 \mbox{\,\,GeV}$).
This may lead to a sufficient strength of
the first order transition to be compatible
with electroweak baryogenesis
for realistic masses of the Higgs
scalar.
For large scalar masses
($M_h > 200 \mbox{\,\,GeV}$) it is not excluded that the barrier 
disappears and the phase transition turns into
an analytical crossover.
We have also estimated lower bounds
on the critical temperature including
nonperturbative effects.
The main uncertainty in the present calculation
concerns the size of nonperturbative
condensates (W-boson condensation), which
determine the behaviour of the
potential for small values of the field.
In view of the extreme sensibility of tunneling
rates to the height of the barrier and the
crucial importance of nonperturbative
phenomena for an estimate of this barrier,
we believe that a quantitative understanding
of the dynamics of the electroweak 
phase transition has to wait for a 
reliable treatment of condensation phenomena
by nonperturbative methods.\bigskip

{\bf Acknowledgement}

\noindent The authors would
like to thank 
W.\,Buchm\"uller and M.\,Shaposhnikov
for fruitful discussions.

%
%
\newpage
\section*{Figure Caption}

\noindent{\bf Fig.\,\,1} Different
approximations to the effective
quartic scalar coupling
for $M_h = 35 \mbox{\ GeV}$ Higgs 
at the 
RG-improved critical temperature.
\vspace{1cm}

\noindent{\bf Fig.\,\,2} 
Renormalization group improved 
effective potential for 
$M_h = 35$ and $80$ GeV
at different temperatures.
\vspace{1cm}

\noindent{\bf Fig.\,\,3} Comparison
of the effective potential
using different $\beta$-functions
for the gauge coupling. 
\vspace{1cm}

\noindent{\bf Fig.\,\,4} Critical 
temperature including condensation
effects as a function of the 
Higgs-mass for different
values of $K$.
\vspace{1cm}

\noindent{\bf Fig.\,\,5} 
Renormalization group improved 
effective potential for 
$M_h = 140$ and $200$ GeV
at different temperatures.

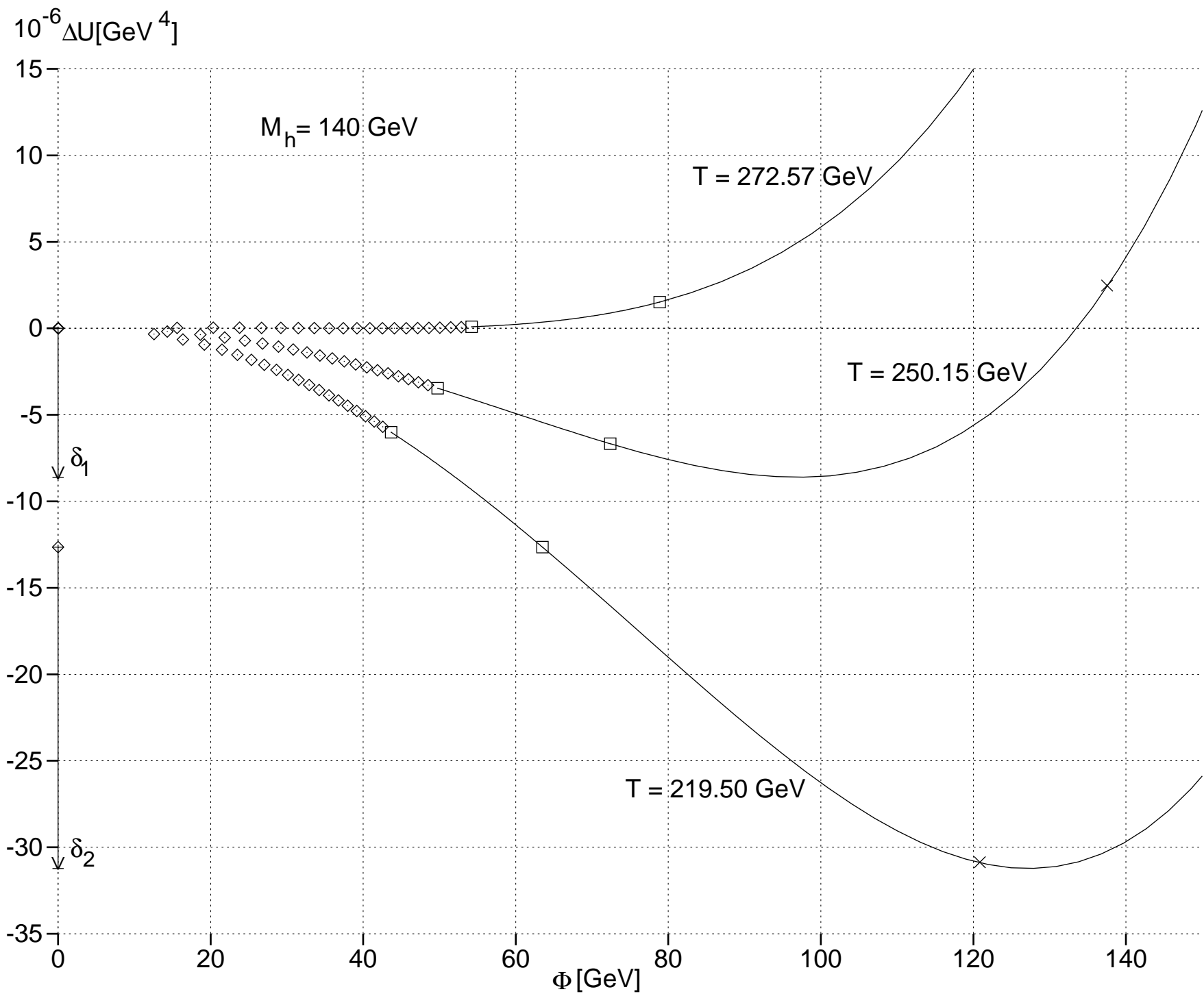
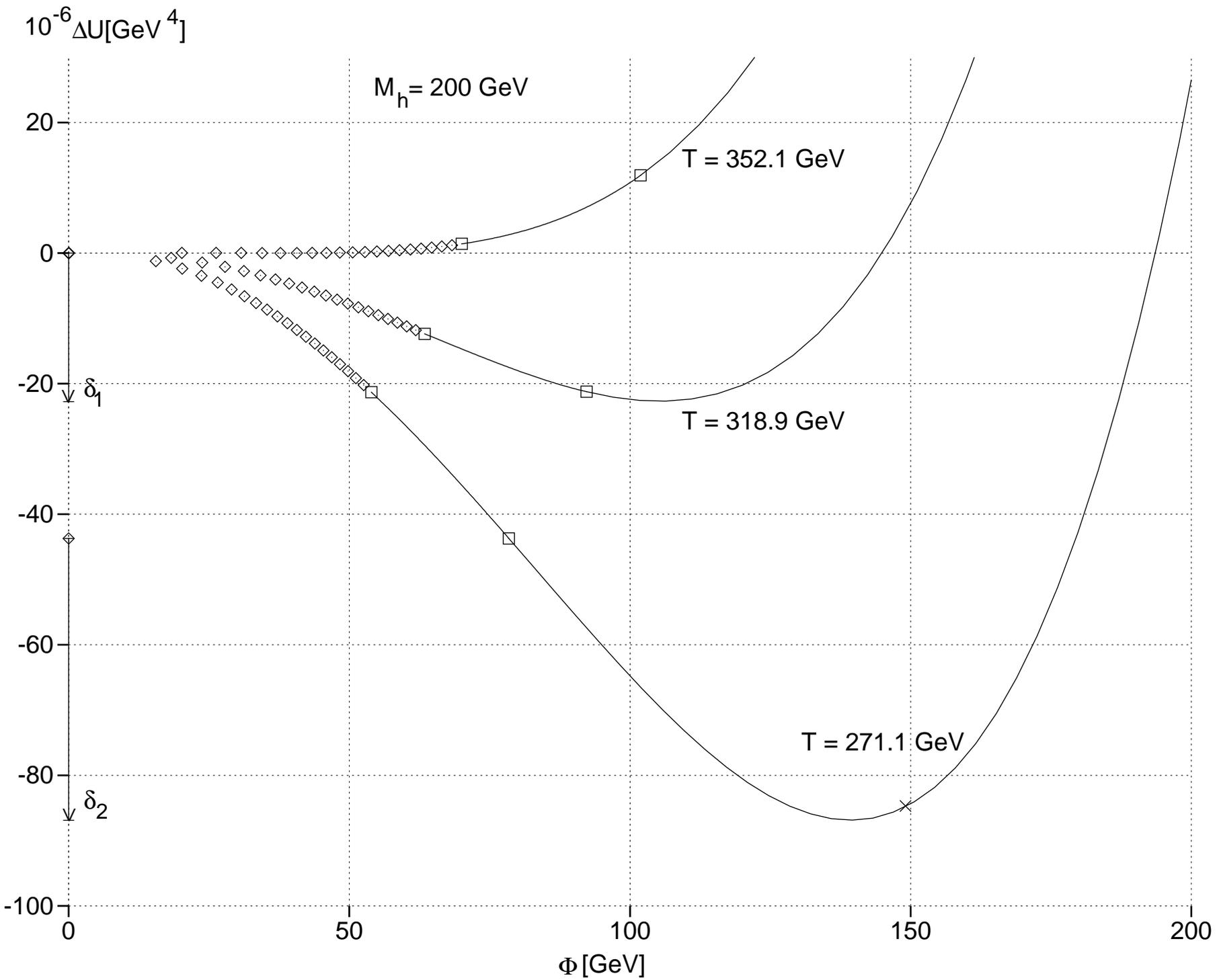
\end{document}